\documentclass[12pt,review,sort&compress]{elsarticle}

\usepackage[utf8]{inputenc}

\usepackage[squaren]{SIunits}
\addunit{\pixel}{px}
\usepackage{amsmath,amssymb}
\usepackage{miller}

\title{Convergent-Beam EMCD: Benefits, Pitfalls, and Applications}

\usepackage[hidelinks]{hyperref}

\begin{document}

\author[ustem,mcmaster]{S.~Löffler\corref{cor1}}
\ead{stefan.loeffler@tuwien.ac.at}

\author[ustem]{W.~Hetaba\fnref{fn1}}

\cortext[cor1]{Corresponding author}

\address[ustem]{University Service Centre for Transmission Electron Microscopy, TU Wien, Vienna, Austria}
\address[mcmaster]{Dept. for Materials Science and Engineering, McMaster University, Hamilton, Ontario, Canada}

\fntext[fn1]{Currently at Fritz-Haber-Institut der Max-Planck-Gesellschaft, Berlin, Germany}


\begin{abstract}
Energy-loss magnetic chiral dichroism (EMCD) is a versatile method for studying magnetic properties on the nanoscale. However, the classical EMCD technique is notorious for its low signal to noise ratio (SNR). Here, we study the theoretical possibilities of using a convergent beam for EMCD.  In particular, we study the influence of detector positioning as well as convergence and collection angles on the detectable EMCD signal. In addition, we analyze the expected SNR and give guidelines for achieving optimal EMCD results.
\end{abstract}

\maketitle

\section{Introduction}

Electron magnetic chiral dichroism (EMCD), the electron microscopic equivalent to X-ray magnetic circular dichroism (XMCD), is a very versatile tool for investigating magnetic materials on the nanometer scale. Ever since its theoretical prediction \cite{U_v96_i_p463} and subsequent realization \cite{N_v441_i_p486}, EMCD has been gaining popularity in many fields, including magnetic nano-engineering and spintronics.

There are, however, two severe limitations with the classical EMCD approach: spatial resolution and signal-to-noise (SNR) ratio. In the classical EMCD approach, one sends a plane wave into a crystal that was tilted into systematic row condition and subsequently measures the inelastically scattered electrons at particular points of the diffraction plane far away from the diffraction spots (see also fig.~\ref{fig:setup}). While plane waves are well-suited for an elegant theoretical treatment, they are not so useful in practice. First of all, from a fundamental point of view, it is impossible to actually create or measure true plane waves, due to the limited extent of the microscope and the apertures, as well as the beam rotation induced by the magnetic lenses \cite{U_v106_i11-12_p1144}. Secondly, from an experimental point of view, a (quasi) plane wave has a very low current density at the sample. Together with the fact that the signal has to be measured off-axis --- where it can be orders of magnitude smaller than on-axis --- with (ideally infinitely) small detectors, this results in a notoriously low SNR. 
Another issue is resolution. When acquiring spectra in diffraction mode, the spatial resolution is usually defined by using a selected area aperture (typically of the order of \unit{100}{\nano\meter}), thereby reducing the signal even further. Alternatively, one can measure in image mode using energy-filtered TEM (EFTEM) \cite{U_v110_i11_p1380,M_v42_i5_p456}. Due to the required energy-slit, this again leads to low intensity, in addition to poor energy resolution.

To overcome these limitations, several approaches have been proposed, ranging from alternative measurement geometries in scanning transmission electron microscopy (STEM) \cite{U_v110_i8_p1038,JoAP_v107_i9_p9,PRB_v82_i_p144418,PRB_v85_i_p134422}, over vortex beams \cite{N_v467_i7313_p301,U_v150_i_p16,PRB_v89_i_p134428}, to the use of aberration correctors to manipulate the phase of the electron beam \cite{PRL_v113_i_p145501}. However, all these methods exhibit very low signal, are typically limited to atomic resolution \cite{PRL_v111_i_p105504,U_v136_i_p81}, and may require changing components of the microscope or operating it under non-standard conditions. Thus, these new methods are not yet applicable for many practical applications.

Here, we analyze another way to improve both the spatial resolution and the SNR at the same time while making use of the original, straight-forward measurement setup: using a convergent beam and finite collection apertures instead of plane waves. While this method has been used experimentally at several occasions to boost the spatial resolution of classical EMCD (see, e.g., \cite{U_v108_i5_p393,U_v108_i5_p433,PRB_v78_i10_p104413,NL_v12_i_p2499,SR_v5_i_p13012}, and it has long been known that large collection apertures can improve the SNR \cite{U_v108_i9_p865}, it is surprising that, to our knowledge, the influence of the convergence angle and the interplay between convergence and collection angle has not been studied extensively from a theoretical point of view before.

In this work, we present simulations that show that convergent beam EMCD is in many ways superior to classical EMCD. In particular, we present simple rules of thumb for how to obtain a substantial improvement of the SNR while at the same time improving the spatial resolution to close to atomic resolution. This is expected to open new avenues for optimizing EMCD measurements in general, but particularly for the characterization of fine grained materials, thin films, as well as the magnetic structure in the vicinity of interfaces and defects. Thus, it is expected to lead to great advances in material science.

\section{Methods}

\begin{figure}
	\includegraphics{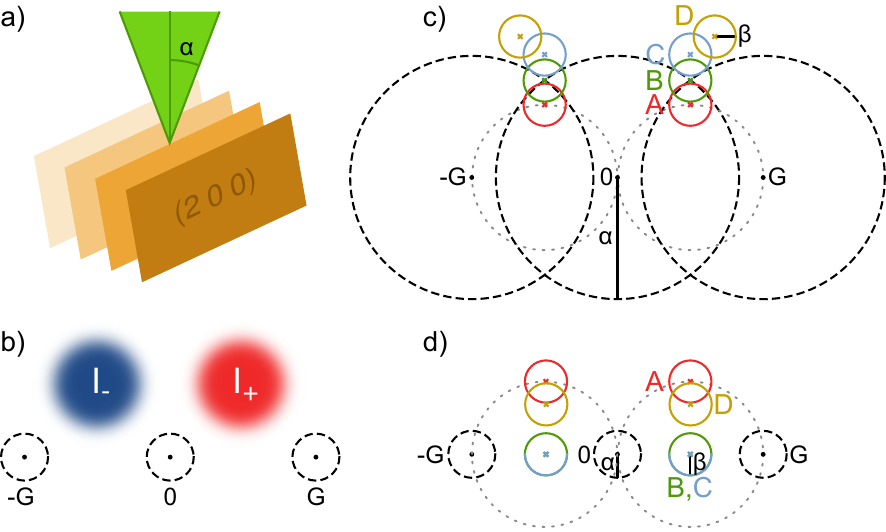}
	\caption{Sketch of the convergent beam setup. (a) The incident beam with convergence semi-angle $\alpha$ is centered on a crystal plane. (b) Sketch of the general positions of the areas with ``positive'' (i.e., higher than non-magnetic) signal $I_+$ and ``negative'' (i.e., lower than non-magnetic) signal $I_-$. (c) Schematic elastic diffraction pattern for large $\alpha$. (d) Schematic elastic diffraction pattern for small $\alpha$. The diffraction spots are labeled 0, G, -G. Diffraction disks are depicted as black dashed lines, the Thales circles are depicted as gray dotted lines. $\alpha$ is the convergence semi-angle, $\beta$ is the collection semi-angle. The four detector positions A--D are described in the text.}
	\label{fig:setup}
\end{figure}

In this work, we present extensive simulations for the model system of a \unit{10}{\nano\meter} thick bcc Fe crystal, tilted \unit{10}{\degree} from the \hkl[0 0 1] zone axis (ZA) to produce a systematic row case including the \hkl(2 0 0) diffraction spot. All simulations were performed using an acceleration voltage of \unit{300}{\kilo\volt} without spherical aberration\footnote{The spherical aberration is not expected to play a major role here, though, as we are working mostly in the diffraction pattern.}. The beam was focused (with varying convergence semi-angle $\alpha$) onto the entry surface of the sample and positioned on an atomic plane. The complete measurement setup is depicted in fig.~\ref{fig:setup}.

The inelastic scattering was performed using the mixed dynamic form factor (MDFF) approach \cite{M_v31_i4_p333,N_v441_i_p486,U_v131_i0_p39}. The MDFF was modeled with an idealized fully spin-polarized cross-density of states \cite{U_v131_i0_p39} and Slater-type orbital wavefunctions \cite{M_v43_i9_p971}, taking into account the dipole allowed transitions $2p \to d$. The elastic scattering both before and after the inelastic scattering were taken into account using the multislice algorithm \cite{Kirkland1998}. A $2048 \times 2048$ grid with $\approx\unit{0.09}{\angstrom\per\pixel}$ was used together with a slice thickness of \unit{1}{\angstrom} and the electrostatic potentials given by Kirkland \cite{Kirkland1998}.

For extracting the relative EMCD effect, one needs to measure the signal strengths at two different positions $I_+, I_-$ and then divide the difference of the two by their average \cite{U_v96_i_p463,U_v108_i3_p277,JoAP_v107_i9_p9}
\begin{equation}
	S = 2 \cdot \frac{I_+ - I_-}{I_+ + I_-} = \frac{\Delta I}{I_0},\ I_0 = \frac{I_+ + I_-}{2}.
\end{equation}
In some cases, only the difference signal $\Delta I$ is used instead of the relative EMCD effect, especially in low-signal/high-noise situations. Therefore, we will also study how to obtain the difference signal in a convergent beam geometry and what SNR can really be achieved that way.

To that end, two different schemes were used. On the one hand, a pointwise comparison of corresponding points on the upper/lower or left/right halves of the diffraction plane was performed to obtain a visual indication of the distribution of the EMCD effect. On the other hand, circular collection apertures (of varying collection semi-angle $\beta$) were centered at four different points of the diffraction plane: (A) on the Thales circle, (B) at the intersection of adjacent elastic diffraction disks\footnote{In case the elastic diffraction disks did not overlap, the apertures were centered on the systematic row}, (C) just outside the elastic diffraction disks such that the collection aperture touched adjacent diffraction disks\footnote{In case such a touching configuration was not possible, the aperture was positioned on the systematic row}, (D) in an ``optimal position'', i.e. at a convergence and collection angle dependent point determined by a downhill simplex optimization algorithm \cite{TCJ_v7_i4_p308} where the maximal EMCD effect can be obtained. All four positions are also depicted in fig.~\ref{fig:setup}.

\section{Results}
\subsection{Position of the EMCD Effect}
\label{sec:position}

\begin{figure*}
\includegraphics[width=\textwidth]{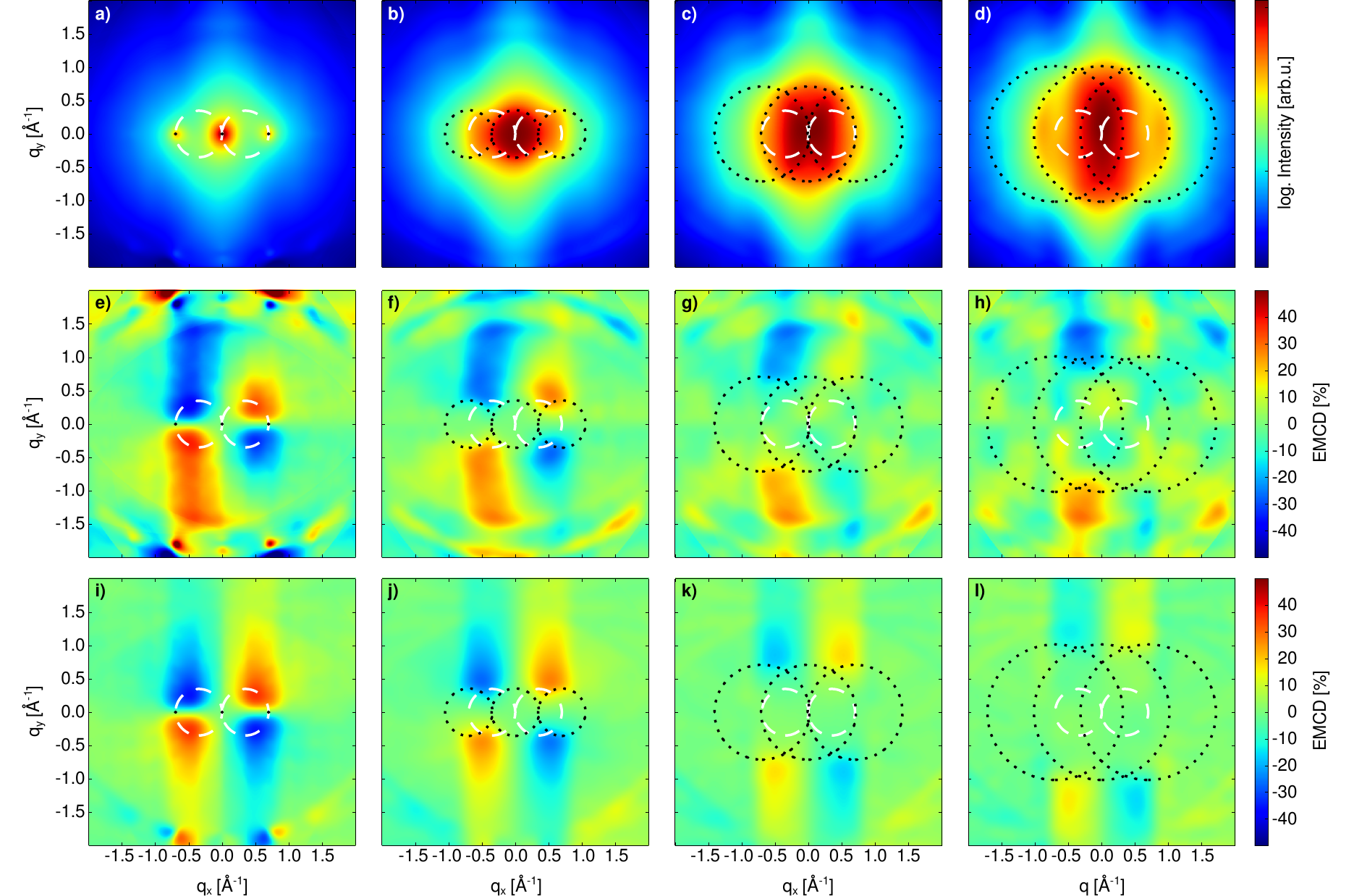}
\caption{Energy-filtered diffraction patterns (a--d), pointwise EMCD maps based on upper/lower halfplane subtraction (e--h) and pointwise EMCD maps based on left/right halfplane subtraction (i--l) for convergence semi-angles of \unit{0}{\milli\radian} (a, e, i), \unit{7}{\milli\radian} (b, f, j), \unit{14}{\milli\radian} (c, g, k), and \unit{20}{\milli\radian} (d, h, l). The black dotted circles indicate the three most intense diffraction disks, whereas the white dashed circles indicate the classical Thales circles. The energy-filtered diffraction patterns are shown in contrast-optimized logarithmic scale.}
\label{fig:EFSAD_example}
\end{figure*}

In order to check the applicability of convergent beam EMCD, it is first necessary to determine where an EMCD effect can be expected in the diffraction plane (if at all). To that end, fig.~\ref{fig:EFSAD_example}a--d show simulated energy filtered diffraction patterns for the Fe L$_3$ edge for different convergence angles. For classical EMCD (i.e., fig.~\ref{fig:EFSAD_example}a), it is well known that there are four areas exhibiting magnetic information, one in each quadrant of the diffraction plane. Therefore, in fig.~\ref{fig:EFSAD_example}e--h, we plotted the EMCD effect calculated pixel by pixel from the difference of the upper and the lower half-plane. Likewise, fig.~\ref{fig:EFSAD_example}i--l, show the EMCD effect calculated pixel by pixel from the difference of the right and the left half-plane.

The first main result from those maps is that with increasing convergence angle, the areas where the EMCD is strong is ``pushed out'' such that it can generally be found close to the rim of the elastic diffraction disks.
This can be explained by considering the relative contributions of the different scattering vectors. Assuming ideal conditions, a point-like detector, and using the dipole approximation \cite{JAP_v103_i7_p7,U_v110_i7_p831,JoAP_v107_i9_p9}, the EMCD difference signal is proportional to 
\begin{equation}
	\int \frac{\vec{q} \times \vec{q}'}{q^2\smash{q'}^2} d^2qd^2q',
\end{equation}
where one has to integrate over \emph{all} combinations of scattering vectors connecting points inside the convergence disks (with radii $\alpha$, see Fig.~\ref{fig:setup}) with the point-like detector. Due to the $1/(q^2\smash{q'}^2)$ dependence, contributions from short scattering vectors are dominant and due to the $\vec{q} \times \vec{q}'$ dependence, contributions are strongest for perpendicular scattering vectors.

In the limit of small convergence angles, only one pair of scattering vectors is possible and the situation reduces to the case of classical EMCD: the perpendicularity requirement suggests that the signal is strongest close to the Thales circle.\footnote{The exact position depends on the characteristic momentum transfer $q_z = q'_z$, as well as the details of the elastic scattering.} For large convergence angles, this explanation no longer holds as then, many combinations of scattering vectors can contribute.

First, we consider detector positions inside the diffraction disks. Without loss of generality, we will assume a detector position inside the 0 diffraction disk. As stated above, the dominant contributions stem from short scattering vectors. For any sufficiently short scattering vector $\vec{q}$ from a point inside the diffraction disk to the detector, the scattering vector $-\vec{q}$ also connects a point inside the diffraction disk to the detector. As the contributions of $(\vec{q}, \vec{q}')$ and $(-\vec{q}, \vec{q}')$ are equal in magnitude but opposite in sign for any scattering vector $\vec{q}'$, all these contributions will average out. This implies that inside the elastic diffraction disks, the EMCD effect will be small. 

Secondly, if the detector is positioned far away from large diffraction disks, neither the perpendicularity constraint nor the shortness requirement can be fulfilled, thus leading to an asymptotically vanishing EMCD effect.

Thirdly, if the detector is positioned close to the intersection of the diffraction disks, there are always pairs of scattering vectors that are short and fulfill the orthogonality requirement, thus yielding an appreciable EMCD effect. 


From fig.~\ref{fig:EFSAD_example}, it is also obvious that the upper/lower difference shows severe left/right differences, particularly for larger scattering angles. While this is of little concern for classical EMCD, where one typically measures at the Thales circle, it does become a large issue for larger convergence angles, where one is forced to measure at larger scattering angles. For the right/left difference maps, however, an upper/lower symmetry generally holds in good approximation except for extremely large scattering angles.

The origin of these different symmetry properties can be found in the tilting of the Ewald sphere with respect to the crystal and the influence of higher order Laue zones (HOLZs), causing an inherent asymmetry of the signal \cite{JoM_v237_i_p465,U_v148_i_p42}. Some artifacts introduced by the HOLZ can be seen particularly well close to the edges of fig.~\ref{fig:EFSAD_example}e\footnote{Note that the figures show only a subset of the total simulated area, so the ``artifacts'' close to the edge are not calculation artifacts but actually coincide with HOLZ reflections consistent for the chosen scattering geometry.}. Due to the asymmetric Ewald sphere and the HOLZ contributions, the intensity in the upper half-plane is slightly lower than the corresponding intensity in the lower half-plane. While this intensity difference is not caused by the spin-polarization of the sample, it can easily be misinterpreted as a ``fake'' EMCD effect. As the setup is symmetric with respect to a right/left mirror operation, the right/left difference maps do not suffer from this effect.

\begin{figure*}
\includegraphics[width=\textwidth]{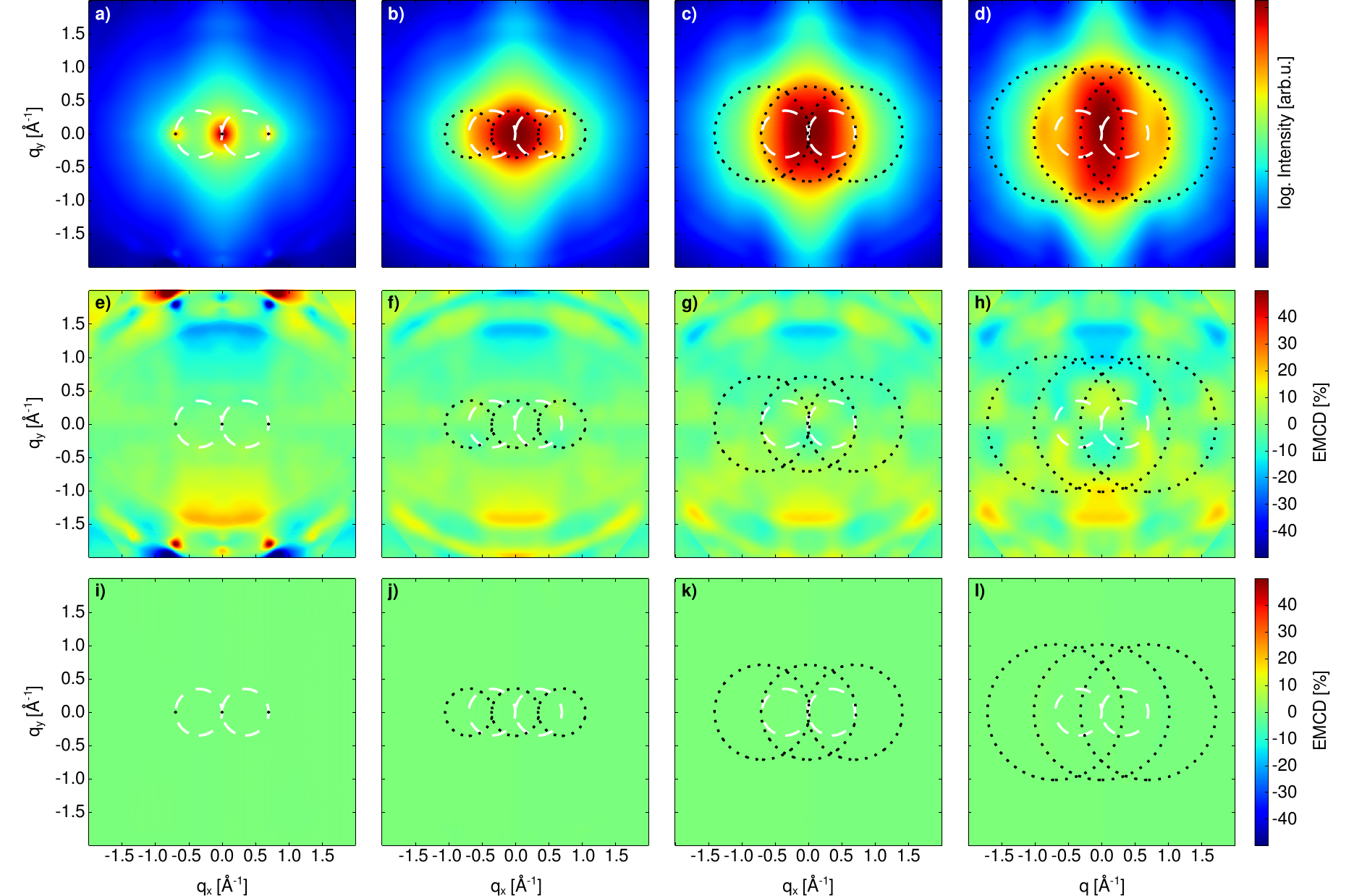}
\caption{Same as Fig.~\ref{fig:EFSAD_example} but for a hypothetical non-magnetic case.}
\label{fig:EFSAD_example_iso}
\end{figure*}

To confirm this interpretation, fig.~\ref{fig:EFSAD_example_iso} shows the same maps, but calculated for a hypothetical ``non-magnetic'' iron where the spin-polarization was forced to zero. Again, the upper/lower difference maps fig.~\ref{fig:EFSAD_example_iso}e--h show a ``fake'' EMCD effect, whereas the right/left difference maps fig.~\ref{fig:EFSAD_example_iso}i--l correctly show no magnetic signal.

Therefore, in the remainder of this work, we use the right/left difference method to extract EMCD signals.

\subsection{EMCD Signal Strength and SNR}

In this section, we will analyze both the achievable signal strengths $S$ and $\Delta I$ as well as the SNR $S/\delta S$ and $\Delta I/\delta\Delta I$ associated with them as a function of convergence and collection angles for the four detector positions A--D defined above. This is conceptionally similar to previous studies that included estimations for the SNR for plane wave illumination \cite{U_v108_i9_p865} and for aberrated probes \cite{PRB_v93_i10_p104420}. To calculate the SNR, we will include the pre-edge background intensity $B$ which does not contribute to the signal but does increase the noise. We will also use the jump ratio defined by
\begin{equation}
	r = \frac{I_0 + B}{B}
\end{equation}
to simplify the equations.

Note that while we will give general formulas that should be applicable to all cases at the beginning of each section, further derivations will be based on the assumption of pure Poissonian shot noise to derive simplified formulas and actual numbers. This neglects other noise sources such as read-out noise and electronic noise (which will be low compared to the shot noise as derived below), or uncertainties introduced by the background subtraction process \cite{Egerton1996}. Nevertheless, the numbers calculated below will give a good rule of thumb for the intensity necessary to obtain a statistically significant EMCD signal.

\subsubsection{Relative EMCD Effect}
\label{sec:EMCD}

\begin{figure*}
	\includegraphics[width=\textwidth]{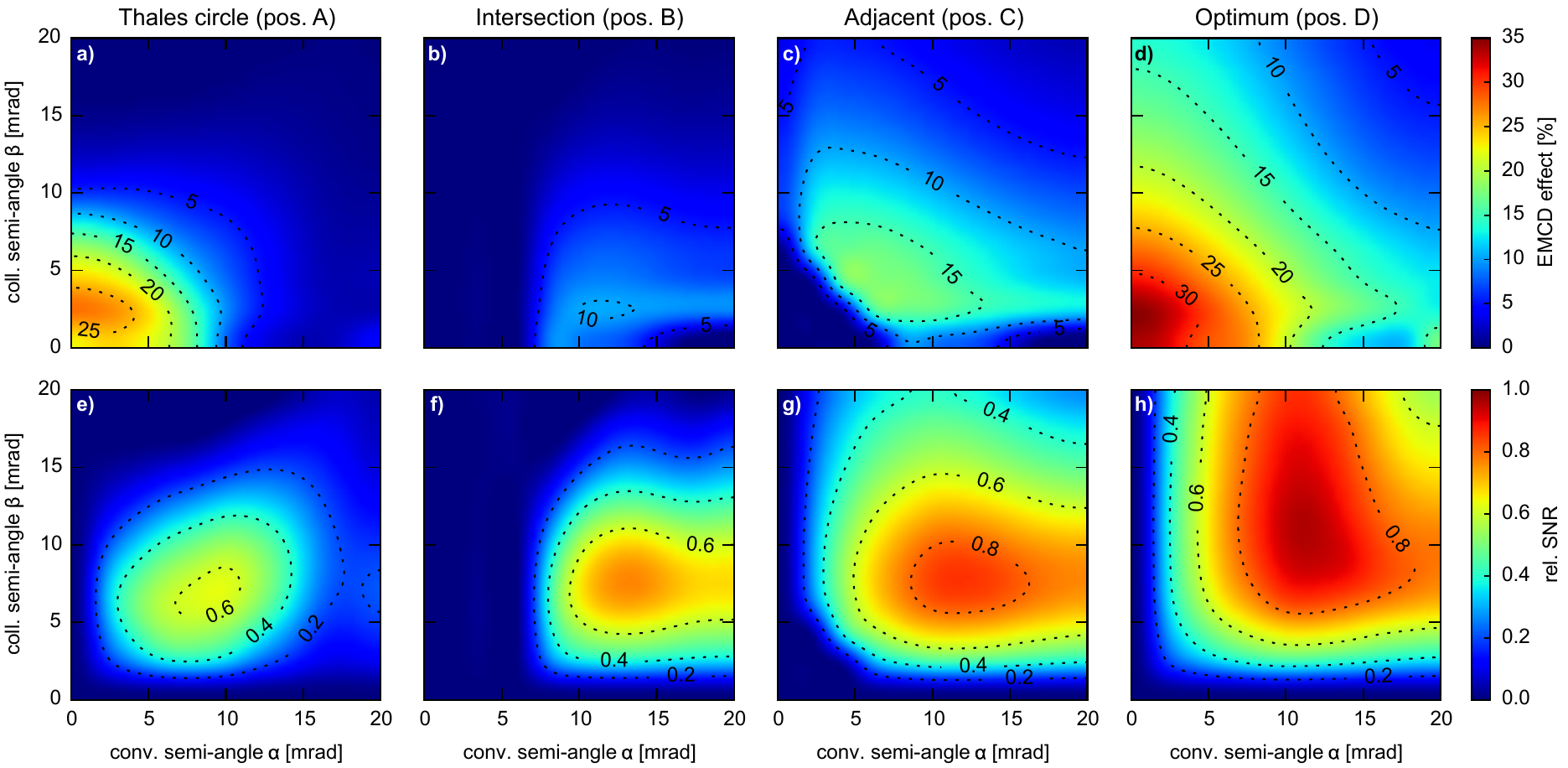}
	\caption{EMCD effect $S$ (a--d) and SNR $S/\delta S$ (e--h) for the four sets of detector positions A--D as a function of convergence and collection semi-angles. The SNR is given for a jump ratio of $r = 2$ in fractions of the maximum SNR.}
	\label{fig:EMCD}
\end{figure*}

Fig.~\ref{fig:EMCD}a--d show the dependence of the EMCD effect on the convergence and collection angles for the four different sets of detector positions A--D defined above. As was already noted in sec.~\ref{sec:position}, placing the detectors on the Thales circle (position A) only gives a large EMCD signal for small convergence and collection angles. For angles larger than the Bragg angle of $\theta_B \approx \unit{6.9}{\milli\radian}$, the signal decreases rapidly as one is then measuring ``inside'' the elastic diffraction disk.

Putting the detectors on the intersection of the elastic diffraction disks (position B) naturally gives no signal for $\alpha < \theta_B$ and then gives a relatively small EMCD signal for small collection angles. This can be understood from the fact that for large collection angles, a significant portion of the collected intensity stems from the areas inside the diffraction disk which does not show a significant EMCD asymmetry.

Putting the detectors adjacent to the elastic diffraction disks (position C) gives medium EMCD effects, but over a large range of convergence and collection angles (apart from the area of $\alpha + \beta < \theta_B$, where the notion of ``adjacent'' does not make sense). In fact, this case is mostly complementary to the Thales circle case.

The final case --- putting the detectors at optimal positions D --- naturally gives the largest EMCD effects, basically combining the ``best of both worlds'': positions A and C. While this case gives the highest EMCD signal by design, it is likely difficult to implement in many applications as it requires extensive simulations with conditions that vary from situation to situation (e.g., with crystal thickness and orientation).

While fig.~\ref{fig:EMCD}a--d shows that convergent beam EMCD works to produce a dichroic signal, it also indicates that the achievable EMCD effect is decreasing somewhat with increasing convergence and collection angle. What was not taken into account so far, however, is the influence of the SNR. 
If shot noise dominates over other noise sources (such as read out noise), $I_\pm$ follows a Poisson distribution. By the central limit theorem, this can be approximated well by a Gaussian distribution with a standard deviation of $\delta I_\pm = \sqrt{I_\pm + B}$ for sufficiently large signal, where $B$ is the background intensity. Then the variance $(\delta S)^2$ of the signal $S$ can be calculated by error propagation to read
\begin{equation}
	(\delta S)^2 = 16 \cdot \frac{(\delta I_+)^2 I_-^2 + I_+^2 (\delta I_-)^2}{(I_+ + I_-)^4}.
\end{equation}
with a SNR of
\begin{equation}
	\frac{S}{\delta S} = \frac{I_+^2 - I_-^2}{2\sqrt{(\delta I_+)^2 I_-^2 + I_+^2 (\delta I_-)^2}}.
\end{equation}
The former can be simplified to
\begin{equation}
	(\delta S)^2 = \frac{16 I_+ I_-}{(I_+ + I_-)^3} + 16 B \cdot \frac{I_+^2 + I_-^2}{(I_+ + I_-)^4}
\end{equation}
By virtue of
\begin{equation}
\begin{aligned}
	I_+ - I_- &= S I_0 \\
	I_+ + I_- &= 2 I_0 \\
	4I_+I_- &= I_0^2 (4 - S^2) \\
	2 (I_+^2 + I_-^2) &= I_0^2 (4 + S^2)
\end{aligned}
\end{equation}
this can also be written as
\begin{equation}
	(\delta S)^2 = \frac{I_0(4-S^2) + B \cdot (4+S^2)}{2I_0^2}.
\end{equation}
Thus, the SNR becomes
\begin{equation}
	\frac{S}{\delta S} = \frac{\sqrt{2} S I_0}{\sqrt{I_0(4-S^2) + B \cdot (4+S^2)}}.
	\label{eq:SNR_EMCD}
\end{equation}
Not surprisingly, the SNR increases with total intensity and EMCD effect and decreases with pre-edge background.

Fig.~\ref{fig:EMCD}e--h depicts the SNR associated with the four sets of detector positions. It clearly confirms the experimental evidence that for very small convergence and/or collection angles, the SNR drops dramatically due to the greatly reduced recorded intensity for a given exposure time. When employing convergent beam EMCD, however, the SNR can easily be increased dramatically. Note that the SNR naturally takes into account the decreasing EMCD effect with larger convergence/collection angles, but the increase in recorded intensity and correspondingly decreasing noise more than compensate for it, giving more reliable and statistically significant results under otherwise identical recording conditions. As a good rule of thumb, one can use a convergence semi-angle slightly larger than $\theta_B$ --- which results in slightly intersecting elastic diffraction disks ---, a collection angle of about $\theta_B$, and position the collection aperture adjacent to the diffraction disks (position C). For the present case studied here, this results in approximately $\unit{80}{\%}$ of the maximal achievable SNR and an EMCD effect of \unit{15}{\%}.

To answer the question of how many counts need to be recorded to achieve a certain statistical significance, one naturally needs to consider the ratio between the elemental edge and the pre-edge background (which increases the noise level but not the signal). Assuming a jump ratio $r$ of
\begin{equation}
	r = \frac{I_0 + B}{B} \ \Leftrightarrow \ B = \frac{I_0}{r - 1} \ \Leftrightarrow \ I_0 + B = I_0 \cdot \frac{r}{r-1},
\end{equation}
the SNR can be rewritten as
\begin{equation}
	\frac{S}{\delta S} = \frac{\sqrt{2 I_0} S}{\sqrt{4-S^2 + \frac{4+S^2}{r-1}}}
	= \frac{S \sqrt{2 (r-1) I_0}}{\sqrt{r(4 - S^2) + 2S^2}}.
\end{equation}
If $B = I_0$, i.e. for a jump ratio of $r=2$, the SNR takes the particularly simple form of
\begin{equation}
	\frac{S}{\delta S} = \frac{S \sqrt{I_0}}{2}.
\end{equation}
To reach a SNR of at least $k$, $I_0$ must be chosen such that
\begin{equation}
	I_0 \ge \frac{k^2}{2S^2} \left( 4-S^2 + \frac{4+S^2}{r-1} \right)
\end{equation}
or, equivalently, that the total intensity fulfills
\begin{equation}
	I_0 + B \ge \frac{k^2r}{2S^2(r-1)} \left( 4-S^2 + \frac{4+S^2}{r-1} \right)
\end{equation}
For the special case of $k = 3$ and $r = 2$, this gives
\begin{equation}
	I_0 + B \ge \frac{72}{S^2}
\end{equation}
i.e., for an expected EMCD effect of \unit{10}{\%}, an intensity of at least 7200 counts needs to be achieved.

\subsubsection{Difference Signal}

In low signal/large noise situations, it is often assumed that the division by $I_0$ renders the relative EMCD effect statistically unstable. In such cases, one might consider looking only at the difference signal $\Delta I$ which, though not being quantifiable in absolute numbers, can still give a qualitative indication of whether a magnetic signal is non-zero and how it changes across the sample. Therefore, here we will also look at the signal strength and the SNR properties of the difference signal $\Delta I$.

\begin{figure*}
	\includegraphics[width=\textwidth]{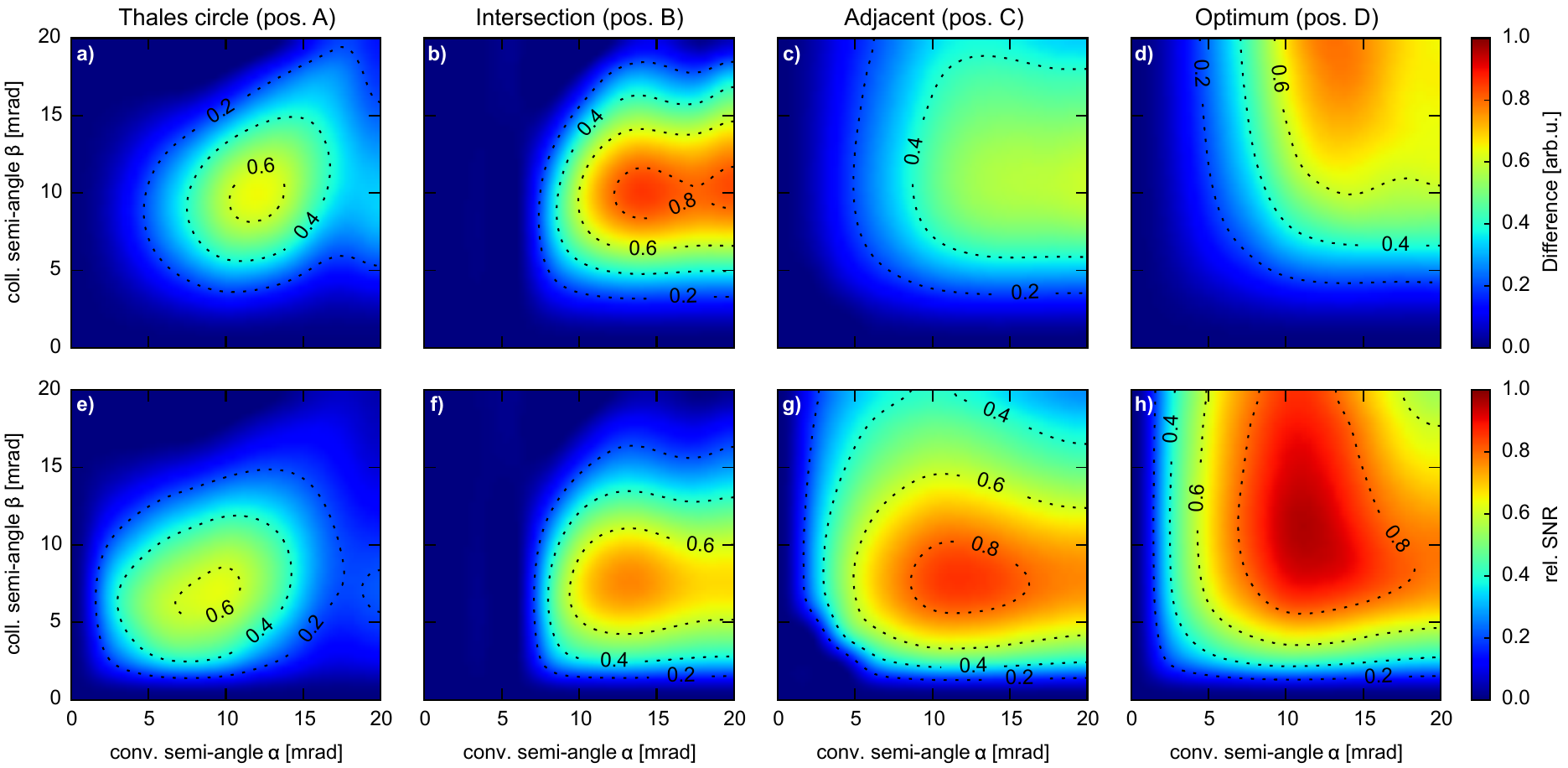}
	\caption{Difference signal $\Delta I$ (a--d) and SNR $\Delta I/\delta \Delta I$ (e--h) for the four sets of detector positions A--D as a function of convergence and collection semi-angles. The SNR is given for a jump ratio of $r = 2$ in fractions of the maximum SNR.}
	\label{fig:Diff}
\end{figure*}

Fig.~\ref{fig:Diff}a--d shows the difference signal dependence on the convergence and collection semi-angles for the four sets of detector positions. Three features immediately catch the eye: first of all, unlike the relative EMCD effect in fig.~\ref{fig:EMCD}, the signal strength tends to increase with increasing convergence and collection angles. Secondly, position B (on the intersection of the diffraction disks) yields higher signal strengths, in stark contrast to the case described in sec.~\ref{sec:EMCD}, where position B exhibited by far the lowest intensity. Thirdly, the region with appreciable signal for position A (on the Thales circle) moved from low convergence/collection angles to medium ones where the Thales circle position is close to the intersection of the diffraction disks. All these effects can easily be understood from the method of signal extraction. While the calculation of the EMCD effect $S$ includes the division by the average signal, the calculation of the difference signal $\Delta I$ does not. Consequently, adding the intensity from points that show little or no asymmetry does not alter the difference signal, while it decreases the EMCD effect. Of course, for both methods, adding points that contribute little to nothing to the signal inherently decreases the SNR, though, and as such should be avoided.

For calculating the SNR, we follow the same steps as in sec.~\ref{sec:EMCD}. Under the same assumptions as above, the variance of the difference signal $\Delta I$ is given by
\begin{equation}
	(\delta \Delta I)^2 = (\delta I_+)^2 + (\delta I_-)^2 = I_+ + I_- + 2B = 2(I_0 + B)
\end{equation}
and the SNR reads
\begin{equation}
	\frac{\Delta I}{\delta \Delta I}
	= \frac{I_+ - I_-}{\sqrt{(\delta I_+)^2 + (\delta I_-)^2}}
	= \frac{S I_0}{\sqrt{2(I_0 + B)}} = \frac{S \sqrt{I_0(r-1)}}{\sqrt{2r}}
	\label{eq:SNR_diff}
\end{equation}
From the third expression, it is obvious that the SNR increases with the EMCD effect and the total intensity while it decreases with increasing background intensity as expected. If $B = I_0$, i.e. for a jump ratio of $r=2$, the SNR takes the same form as $S / \delta S$, i.e.,
\begin{equation}
	\frac{\Delta I}{\delta \Delta I} = \frac{\Delta I}{2\sqrt{I_0}} = \frac{S \sqrt{I_0}}{2}.
\end{equation}
To reach an SNR of $k$, one needs to achieve a total intensity of
\begin{equation}
	I_0 + B \ge \frac{2r^2k^2}{S^2(r-1)^2}
\end{equation}
counts.

\section{Discussion}

\subsection{Relative EMCD Effect vs. Difference Signal}

As mentioned above, it is often assumed that the use of the difference signal is beneficial particularly when the signal is weak. This notion presumably comes from the fact that an increase of the exposure time or the incident beam current actually increases the difference signal but does not change the relative EMCD effect. While it can be argued that this complicates post-processing by requiring additional normalizing by the incident dose --- which is inherently included in a way in the relative EMCD signal ---, such a discussion misses the most important point: the SNR. No matter how large or small the EMCD signal itself is, the real question is if it is detectable under given conditions. If a condition A yields a SNR that is good enough to detect a small signal, it is still preferable over a condition B which gives a larger signal but also a bad SNR resulting in an EMCD signal indistinguishable from the noise. Therefore, the crucial aspect is really SNR, and not total signal strength.

A comparison of eq.~\ref{eq:SNR_EMCD} and eq.~\ref{eq:SNR_diff} gives
\begin{align}
	\frac{\sqrt{2} S I_0}{\sqrt{I_0(4-S^2) + B \cdot (4+S^2)}} &< \frac{S I_0}{\sqrt{2(I_0 + B)}} \\
	4(I_0 + B) &< I_0(4-S^2) + B \cdot (4+S^2) \\
	0 &< S^2 \cdot (B-I_0)
\end{align}
Therefore, only for $B>I_O \Leftrightarrow r < 2$, i.e. for thick specimens, using the difference signal is actually better than using the relative EMCD effect. However, thick specimens typically yield a low overall EMCD effect owing to oscillations  and sign reversal caused by the elastic scattering and pendellösung \cite{PRB_v75_i21_p214425,U_v110_i7_p831}. Therefore, the relative EMCD signal should generally preferred over the difference signal.

\subsection{Beam Position Dependence}

In this section, we investigate the dependence of the convergent beam EMCD signal on the beam position. For small convergence and collection angles, one can expect that the EMCD signal is largely independent of the beam position due to the large illuminated area and, consequently, the low spatial resolution. For convergence and collection semi-angles significantly larger than the Bragg angle, however, one can expect a position-dependence. To study this effect, we also performed calculations with the beam displaced by half a lattice plane distance so that it was positioned directly in-between adjacent lattice planes.

\begin{figure}
	\includegraphics[width=90mm]{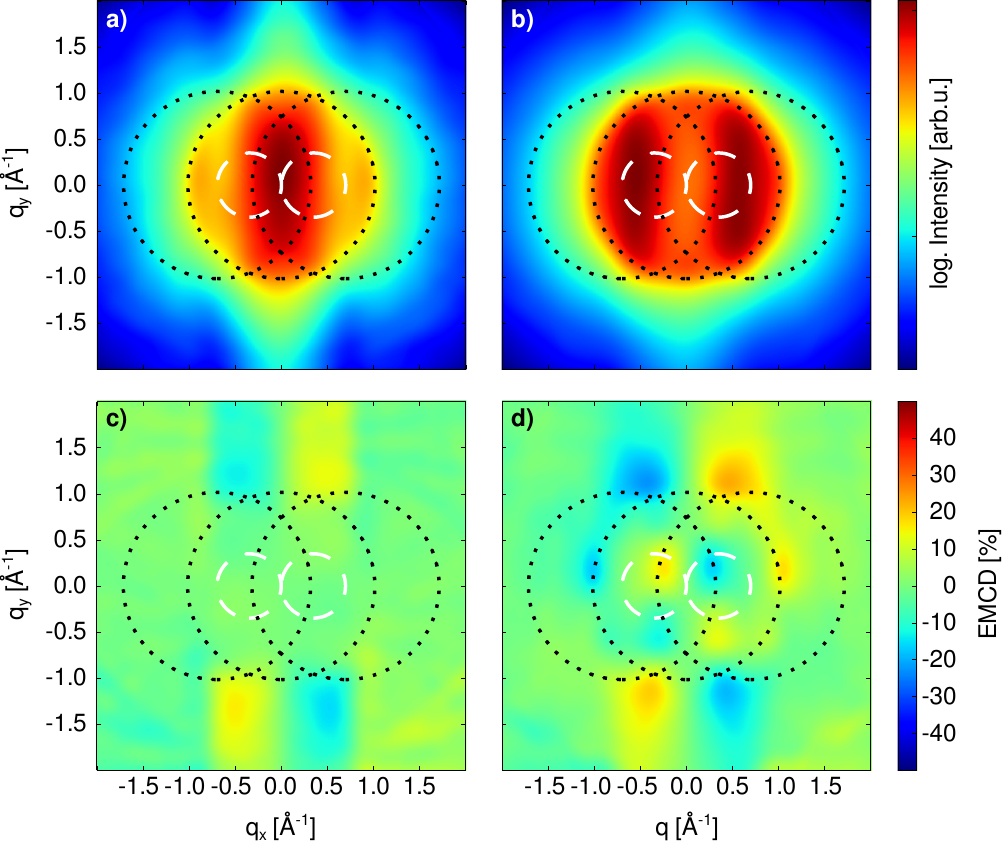}
	\caption{Energy-filtered diffraction patterns (a, b) and pointwise EMCD maps based on left/right halfplane subtraction (c, d) for on-plane (a, c) and off-plane (b, d) beam positions. The convergence semi-angle is \unit{20}{\milli\radian}. The black dotted circles indicate the three most intense diffraction disks, whereas the white dashed circles indicate the classical Thales circles. The energy-filtered diffraction patterns are shown in contrast-optimized logarithmic scale.}
	\label{fig:EFSAD_8px}
\end{figure}

Fig.~\ref{fig:EFSAD_8px} compares the energy-filtered diffraction patterns and point-wise EMCD effects for on-plane and off-plane beam positions for a large convergence angle. While there are obvious differences, it is remarkable that an EMCD effect with the same sign can be found at the same positions adjacent to the diffraction disks. In fact, for the off-plane condition, the EMCD effect is stronger than for the on-plane condition. Qualitatively, this can be understood from the fact that the inelastic scattering kernels contributing to EMCD have the same shape as electron vortex beams: an azimuthal phase ramp combined with a donut-shaped intensity distribution \cite{U_v131_i0_p39,U_v109_i7_p781,APL_v99_i20_p203109}. Thus, the highest probability for exciting a transition that contributes to the EMCD signal with a very small probe is actually not on the atomic nuclei, but in the area surrounding them.\footnote{This can also be understood from the fact that the initial p-states contributing to the L-edge have vanishing probability density at the position of the nucleus.} Of course, the question of how much which atom contributes to the EMCD effect depends crucially on how the incident and outgoing electron beams channel through the crystal \cite{MaM_v18_i_p711,ACA_v68_i_p443}. However, a full quantitative description of the resulting thickness dependence is beyond the scope of this work.

\begin{figure*}
	\includegraphics[width=\textwidth]{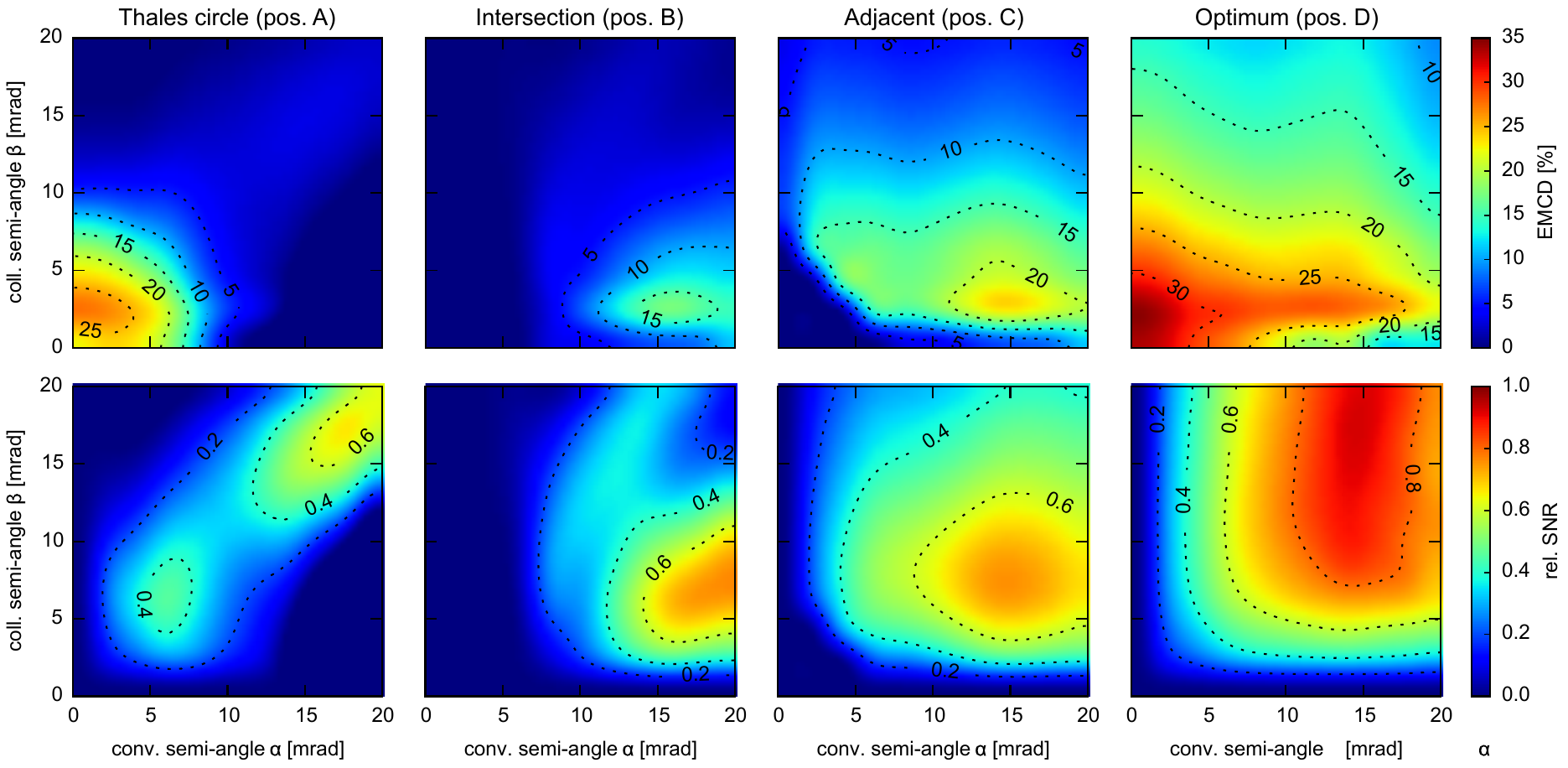}
	\caption{EMCD effect $S$ (a--d) and SNR $S/\delta S$ (e--h) for the four sets of detector positions A--D as a function of convergence and collection semi-angles for a beam position in-between atomic planes. The SNR is given for a jump ratio of $r = 2$ in fractions of the maximum SNR.}
	\label{fig:EMCD_8px}
\end{figure*}

Fig.~\ref{fig:EMCD_8px} shows the convergence and collection semi-angle dependence of the EMCD signal for a probe beam positioned between atomic planes, together with the corresponding SNR. It is not surprising that, qualitatively, it looks similar to the on-plane case depicted in fig.~\ref{fig:EMCD}. In particular for small convergence and collection semi-angles, the maps are identical, as is to be expected. Perhaps the most noticeable difference is the different large-angle behavior of the SNR for positions A and B. This can be understood from the fact that both positions pick up intensity from inside the area of the elastic diffraction disks. This intensity is strongly influenced by the local potential the beam traverses \cite{NC_v5_i_p5653}, and, hence, strongly dependent on the beam position, as is shown in figs.~\ref{fig:EFSAD_8px}a, b. If the collection aperture is placed outside the area of the elastic diffraction disks, however, as is the case for positions C and D, the off-plane signal becomes remarkably similar to the on-plane signal, with the above-mentioned enhancement for large convergence angles.

\section{Concluding Remarks}

In this work, we have explored the possibility of convergent-beam EMCD. We found that this method should not only give a similar EMCD signal as the classical, parallel beam EMCD method, but in fact is expected to have superior SNR characteristics. As a rule of thumb, choosing a convergence semi-angle slightly larger than the Bragg angle, a collection angle close to the Bragg angle, positioning the collection aperture just outside the elastic diffraction disks, and using an exposure time giving more than 7200 counts at the edge under investigation should give close to optimal results.\footnote{The exact value will depend on the peak-to-background ratio and the expected EMCD effect, which, in turn, will depend on the sample material, thickness, and orientation as well as the scattering geometry.}

Of course, further work is necessary, e.g., to adapt the EMCD sum rules \cite{PRB_v76_i6_p60408} to the convergent beam case and to characterize the thickness-dependence of convergent-beam EMCD. However, especially the improvements in SNR, as well as in spatial resolution, open exciting new possibilities for EMCD that may soon lead to an even broader applicability of this exciting technique for material science.

\section*{Funding}
This work was supported by the Austrian Science Fund (FWF) [grant number J3732-N27].

\section*{Acknowledgements}

The authors gratefully acknowledge access to the USTEM computational facilities, as well as fruitful discussions with Peter Schattschneider.

\section*{Bibliography}
\bibliography{papers,mypubs}

\begin{thebibliography}{10}
\expandafter\ifx\csname url\endcsname\relax
  \def\url#1{\texttt{#1}}\fi
\expandafter\ifx\csname urlprefix\endcsname\relax\def\urlprefix{URL }\fi
\expandafter\ifx\csname href\endcsname\relax
  \def\href#1#2{#2} \def\path#1{#1}\fi

\bibitem{U_v96_i_p463}
C.~Hébert, P.~Schattschneider, A proposal for dichroic experiments in the
  electron microscope, Ultramicroscopy 96~(3-4) (2003) 463 -- 468.
\newblock \href {http://dx.doi.org/10.1016/S0304-3991(03)00108-6}
  {\path{doi:10.1016/S0304-3991(03)00108-6}}.

\bibitem{N_v441_i_p486}
P.~Schattschneider, S.~Rubino, C.~Hebert, J.~Rusz, J.~Kunes, P.~Nov{\'a}k,
  E.~Carlino, M.~Fabrizioli, G.~Panaccione, G.~Rossi, Detection of magnetic
  circular dichroism using a transmission electron microscope, Nat. 441 (2006)
  486--488.
\newblock \href {http://dx.doi.org/10.1038/nature04778}
  {\path{doi:10.1038/nature04778}}.

\bibitem{U_v106_i11-12_p1144}
D.~Eyidi, C.~Hébert, P.~Schattschneider, Short note on parallel illumination
  in the tem, Ultramicroscopy 106~(11-12) (2006) 1144 -- 1149.
\newblock \href {http://dx.doi.org/10.1016/j.ultramic.2006.04.029}
  {\path{doi:10.1016/j.ultramic.2006.04.029}}.

\bibitem{U_v110_i11_p1380}
H.~Lidbaum, J.~Rusz, S.~Rubino, A.~Liebig, B.~Hjörvarsson, P.~M. Oppeneer,
  O.~Eriksson, K.~Leifer, Reciprocal and real space maps for emcd experiments,
  Ultramicroscopy 110~(11) (2010) 1380 -- 1389.
\newblock \href {http://dx.doi.org/10.1016/j.ultramic.2010.07.004}
  {\path{doi:10.1016/j.ultramic.2010.07.004}}.

\bibitem{M_v42_i5_p456}
M.~Stöger-Pollach, C.~Treiber, G.~Resch, D.~Keays, I.~Ennen, Emcd real space
  maps of magnetospirillum magnetotacticum, Micron 42~(5) (2011) 456 -- 460.
\newblock \href {http://dx.doi.org/10.1016/j.micron.2011.01.003}
  {\path{doi:10.1016/j.micron.2011.01.003}}.

\bibitem{U_v110_i8_p1038}
P.~Schattschneider, I.~Ennen, M.~Stöger-Pollach, J.~Verbeeck, V.~Mauchamp,
  M.~Jaouen, Real space maps of magnetic moments on the atomic scale: Theory
  and feasibility, Ultramicroscopy 110~(8) (2010) 1038 -- 1041.
\newblock \href {http://dx.doi.org/10.1016/j.ultramic.2009.11.020}
  {\path{doi:10.1016/j.ultramic.2009.11.020}}.

\bibitem{JoAP_v107_i9_p9}
P.~Schattschneider, I.~Ennen, S.~Löffler, M.~Stöger-Pollach, J.~Verbeeck,
  \href{http://link.aip.org/link/?JAP/107/09D311/1}{Circular dichroism in the
  electron microscope: Progress and applications (invited)}, J. Appl. Phys.
  107~(9) (2010) 09D311.
\newblock \href {http://dx.doi.org/10.1063/1.3365517}
  {\path{doi:10.1063/1.3365517}}.
\newline\urlprefix\url{http://link.aip.org/link/?JAP/107/09D311/1}

\bibitem{PRB_v82_i_p144418}
P.~Schattschneider, J.~Verbeeck, V.~Mauchamp, M.~Jaouen, A.-L. Hamon,
  Real-space simulations of spin-polarized electronic transitions in iron,
  Phys. Rev. B 82 (2010) 144418.
\newblock \href {http://dx.doi.org/10.1103/PhysRevB.82.144418}
  {\path{doi:10.1103/PhysRevB.82.144418}}.

\bibitem{PRB_v85_i_p134422}
P.~Schattschneider, B.~Schaffer, I.~Ennen, J.~Verbeeck, Mapping spin-polarized
  transitions with atomic resolution, Phys. Rev. B 85 (2012) 134422.
\newblock \href {http://dx.doi.org/10.1103/PhysRevB.85.134422}
  {\path{doi:10.1103/PhysRevB.85.134422}}.

\bibitem{N_v467_i7313_p301}
J.~Verbeeck, H.~Tian, P.~Schattschneider, Production and application of
  electron vortex beams, Nat. 467~(7313) (2010) 301--304.
\newblock \href {http://dx.doi.org/10.1038/nature09366}
  {\path{doi:10.1038/nature09366}}.

\bibitem{U_v150_i_p16}
D.~Pohl, S.~Schneider, J.~Rusz, B.~Rellinghaus,
  \href{http://www.sciencedirect.com/science/article/pii/S0304399114002411}{Electron
  vortex beams prepared by a spiral aperture with the goal to measure {EMCD} on
  ferromagnetic films via {STEM}}, Ultramicroscopy 150 (2015) 16 -- 22.
\newblock \href
  {http://dx.doi.org/http://dx.doi.org/10.1016/j.ultramic.2014.11.025}
  {\path{doi:http://dx.doi.org/10.1016/j.ultramic.2014.11.025}}.
\newline\urlprefix\url{http://www.sciencedirect.com/science/article/pii/S0304399114002411}

\bibitem{PRB_v89_i_p134428}
J.~Rusz, S.~Bhowmick, M.~Eriksson, N.~Karlsson,
  \href{http://link.aps.org/doi/10.1103/PhysRevB.89.134428}{Scattering of
  electron vortex beams on a magnetic crystal: Towards atomic-resolution
  magnetic measurements}, Phys. Rev. B 89 (2014) 134428.
\newblock \href {http://dx.doi.org/10.1103/PhysRevB.89.134428}
  {\path{doi:10.1103/PhysRevB.89.134428}}.
\newline\urlprefix\url{http://link.aps.org/doi/10.1103/PhysRevB.89.134428}

\bibitem{PRL_v113_i_p145501}
J.~Rusz, J.-C. Idrobo, S.~Bhowmick,
  \href{http://link.aps.org/doi/10.1103/PhysRevLett.113.145501}{Achieving
  atomic resolution magnetic dichroism by controlling the phase symmetry of an
  electron probe}, Phys. Rev. Lett. 113 (2014) 145501.
\newblock \href {http://dx.doi.org/10.1103/PhysRevLett.113.145501}
  {\path{doi:10.1103/PhysRevLett.113.145501}}.
\newline\urlprefix\url{http://link.aps.org/doi/10.1103/PhysRevLett.113.145501}

\bibitem{PRL_v111_i_p105504}
J.~Rusz, S.~Bhowmick,
  \href{http://link.aps.org/doi/10.1103/PhysRevLett.111.105504}{Boundaries for
  efficient use of electron vortex beams to measure magnetic properties}, Phys.
  Rev. Lett. 111 (2013) 105504.
\newblock \href {http://dx.doi.org/10.1103/PhysRevLett.111.105504}
  {\path{doi:10.1103/PhysRevLett.111.105504}}.
\newline\urlprefix\url{http://link.aps.org/doi/10.1103/PhysRevLett.111.105504}

\bibitem{U_v136_i_p81}
P.~Schattschneider, S.~Löffler, M.~Stöger-Pollach, J.~Verbeeck, Is magnetic
  chiral dichroism feasible with electron vortices?, Ultramicroscopy 136 (2014)
  81--85.
\newblock \href {http://arxiv.org/abs/1304.7976} {\path{arXiv:1304.7976}},
  \href {http://dx.doi.org/10.1016/j.ultramic.2013.07.012}
  {\path{doi:10.1016/j.ultramic.2013.07.012}}.

\bibitem{U_v108_i5_p393}
B.~Warot-Fonrose, F.~Houdellier, M.~Hÿtch, L.~Calmels, V.~Serin, E.~Snoeck,
  Mapping inelastic intensities in diffraction patterns of magnetic samples
  using the energy spectrum imaging technique, Ultramicroscopy 108~(5) (2008)
  393 -- 398.
\newblock \href {http://dx.doi.org/10.1016/j.ultramic.2007.05.013}
  {\path{doi:10.1016/j.ultramic.2007.05.013}}.

\bibitem{U_v108_i5_p433}
P.~Schattschneider, C.~H\'{e}bert, S.~Rubino, M.~St{\"o}ger-Pollach, J.~Rusz,
  P.~Nov\'{a}k, {Magnetic circular dichroism in EELS: Towards 10 nm
  resolution}, Ultramicroscopy 108~(5) (2008) 433 -- 438.
\newblock \href {http://dx.doi.org/10.1016/j.ultramic.2007.07.002}
  {\path{doi:10.1016/j.ultramic.2007.07.002}}.

\bibitem{PRB_v78_i10_p104413}
P.~Schattschneider, M.~St\"{o}ger-Pollach, S.~Rubino, M.~Sperl, C.~Hurm,
  J.~Zweck, J.~Rusz, Detection of magnetic circular dichroism on the
  two-nanometer scale, Phys. Rev. B 78~(10) (2008) 104413.
\newblock \href {http://dx.doi.org/10.1103/PhysRevB.78.104413}
  {\path{doi:10.1103/PhysRevB.78.104413}}.

\bibitem{NL_v12_i_p2499}
J.~Salafranca, J.~Gazquez, N.~P{\'{e}}rez, A.~Labarta, S.~T. Pantelides, S.~J.
  Pennycook, X.~Batlle, M.~Varela,
  \href{http://dx.doi.org/10.1021/nl300665z}{Surfactant organic molecules
  restore magnetism in metal-oxide nanoparticle surfaces}, Nano Letters 12~(5)
  (2012) 2499--2503.
\newblock \href {http://dx.doi.org/10.1021/nl300665z}
  {\path{doi:10.1021/nl300665z}}.
\newline\urlprefix\url{http://dx.doi.org/10.1021/nl300665z}

\bibitem{SR_v5_i_p13012}
T.~Thersleff, J.~Rusz, S.~Rubino, B.~Hjörvarsson, Y.~Ito, N.~J. Zaluzec,
  K.~Leifer, \href{http://dx.doi.org/10.1038/srep13012}{Quantitative analysis
  of magnetic spin and orbital moments from an oxidized iron (1 1 0) surface
  using electron magnetic circular dichroism}, Sci. Rep. 5 (2015) 13012.
\newblock \href {http://dx.doi.org/10.1038/srep13012}
  {\path{doi:10.1038/srep13012}}.
\newline\urlprefix\url{http://dx.doi.org/10.1038/srep13012}

\bibitem{U_v108_i9_p865}
J.~Verbeeck, C.~Hébert, S.~Rubino, P.~Novák, J.~Rusz, F.~Houdellier,
  C.~Gatel, P.~Schattschneider, {Optimal aperture sizes and positions for EMCD
  experiments}, Ultramicroscopy 108~(9) (2008) 865 -- 872.
\newblock \href {http://dx.doi.org/10.1016/j.ultramic.2008.02.007}
  {\path{doi:10.1016/j.ultramic.2008.02.007}}.

\bibitem{M_v31_i4_p333}
P.~Schattschneider, M.~Nelhiebel, H.~Souchay, B.~Jouffrey, The physical
  significance of the mixed dynamic form factor, Micron 31~(4) (2000) 333 --
  345.
\newblock \href {http://dx.doi.org/10.1016/S0968-4328(99)00112-2}
  {\path{doi:10.1016/S0968-4328(99)00112-2}}.

\bibitem{U_v131_i0_p39}
S.~Löffler, V.~Motsch, P.~Schattschneider,
  \href{http://www.sciencedirect.com/science/article/pii/S0304399113000910}{A
  pure state decomposition approach of the mixed dynamic form factor for
  mapping atomic orbitals}, Ultramicroscopy 131 (2013) 39 -- 45.
\newblock \href {http://arxiv.org/abs/1210.2947} {\path{arXiv:1210.2947}},
  \href {http://dx.doi.org/10.1016/j.ultramic.2013.03.021}
  {\path{doi:10.1016/j.ultramic.2013.03.021}}.
\newline\urlprefix\url{http://www.sciencedirect.com/science/article/pii/S0304399113000910}

\bibitem{M_v43_i9_p971}
S.~Löffler, P.~Schattschneider,
  \href{http://www.sciencedirect.com/science/article/pii/S0968432812001047}{Transition
  probability functions for applications of inelastic electron scattering},
  Micron 43~(9) (2012) 971 -- 977.
\newblock \href {http://arxiv.org/abs/1112.5607} {\path{arXiv:1112.5607}},
  \href {http://dx.doi.org/10.1016/j.micron.2012.03.020}
  {\path{doi:10.1016/j.micron.2012.03.020}}.
\newline\urlprefix\url{http://www.sciencedirect.com/science/article/pii/S0968432812001047}

\bibitem{Kirkland1998}
E.~J. Kirkland,
  \href{http://aleph.ub.tuwien.ac.at/F/?func=direct&doc_number=000274136&local_base=TUW01}{Advanced
  computing in electron microscopy}, Plenum Press, 1998.
\newline\urlprefix\url{http://aleph.ub.tuwien.ac.at/F/?func=direct&doc_number=000274136&local_base=TUW01}

\bibitem{U_v108_i3_p277}
C.~Hébert, P.~Schattschneider, S.~Rubino, P.~Novak, J.~Rusz,
  M.~Stöger-Pollach, Magnetic circular dichroism in electron energy loss
  spectrometry, Ultramicroscopy 108~(3) (2008) 277 -- 284.
\newblock \href {http://dx.doi.org/10.1016/j.ultramic.2007.07.011}
  {\path{doi:10.1016/j.ultramic.2007.07.011}}.

\bibitem{TCJ_v7_i4_p308}
J.~A. Nelder, R.~Mead, \href{http://dx.doi.org/10.1093/comjnl/7.4.308}{A
  simplex method for function minimization}, The Computer Journal 7~(4) (1965)
  308–313.
\newblock \href {http://dx.doi.org/10.1093/comjnl/7.4.308}
  {\path{doi:10.1093/comjnl/7.4.308}}.
\newline\urlprefix\url{http://dx.doi.org/10.1093/comjnl/7.4.308}

\bibitem{JAP_v103_i7_p7}
P.~Schattschneider, S.~Rubino, M.~Stöger-Pollach, C.~H\'{e}bert, J.~Rusz,
  L.~Calmels, E.~Snoeck, Energy loss magnetic chiral dichroism: A new technique
  for the study of magnetic properties in the electron microscope (invited), J.
  Appl. Phys. 103~(7) (2008) 07D931.
\newblock \href {http://dx.doi.org/10.1063/1.2836680}
  {\path{doi:10.1063/1.2836680}}.

\bibitem{U_v110_i7_p831}
S.~Löffler, P.~Schattschneider, A software package for the simulation of
  energy-loss magnetic chiral dichroism, Ultramicroscopy 110~(7) (2010)
  831--835.
\newblock \href {http://dx.doi.org/10.1016/j.ultramic.2010.02.044}
  {\path{doi:10.1016/j.ultramic.2010.02.044}}.

\bibitem{JoM_v237_i_p465}
J.~RUSZ, P.~OPPENEER, H.~LIDBAUM, S.~RUBINO, K.~LEIFER,
  \href{http://dx.doi.org/10.1111/j.1365-2818.2009.03295.x}{Asymmetry of the
  two-beam geometry in {EMCD} experiments}, Journal of Microscopy 237~(3)
  (2010) 465--468.
\newblock \href {http://dx.doi.org/10.1111/j.1365-2818.2009.03295.x}
  {\path{doi:10.1111/j.1365-2818.2009.03295.x}}.
\newline\urlprefix\url{http://dx.doi.org/10.1111/j.1365-2818.2009.03295.x}

\bibitem{U_v148_i_p42}
D.~Song, Z.~Wang, J.~Zhu, Effect of the asymmetry of dynamical electron
  diffraction on intensity of acquired {EMC}d signals, Ultramicroscopy 148
  (2015) 42--51.
\newblock \href {http://dx.doi.org/10.1016/j.ultramic.2014.08.012}
  {\path{doi:10.1016/j.ultramic.2014.08.012}}.

\bibitem{PRB_v93_i10_p104420}
J.~Rusz, J.~C. Idrobo,
  \href{http://link.aps.org/doi/10.1103/PhysRevB.93.104420}{Aberrated electron
  probes for magnetic spectroscopy with atomic resolution: Theory and practical
  aspects}, Phys. Rev. B 93 (2016) 104420.
\newblock \href {http://dx.doi.org/10.1103/PhysRevB.93.104420}
  {\path{doi:10.1103/PhysRevB.93.104420}}.
\newline\urlprefix\url{http://link.aps.org/doi/10.1103/PhysRevB.93.104420}

\bibitem{Egerton1996}
R.~F. Egerton, Electron Energy-Loss Spectroscopy in the Electron Microscope,
  2nd Edition, Plenum Press, New York, 1996.

\bibitem{PRB_v75_i21_p214425}
J.~Rusz, S.~Rubino, P.~Schattschneider, First-principles theory of chiral
  dichroism in electron microscopy applied to 3d ferromagnets, Phys. Rev. B
  75~(21) (2007) 214425.
\newblock \href {http://dx.doi.org/10.1103/PhysRevB.75.214425}
  {\path{doi:10.1103/PhysRevB.75.214425}}.

\bibitem{U_v109_i7_p781}
P.~Schattschneider, J.~Verbeeck, A.~Hamon, Real space maps of atomic
  transitions, Ultramicroscopy 109~(7) (2009) 781 -- 787.
\newblock \href {http://dx.doi.org/10.1016/j.ultramic.2009.01.016}
  {\path{doi:10.1016/j.ultramic.2009.01.016}}.

\bibitem{APL_v99_i20_p203109}
J.~Verbeeck, P.~Schattschneider, S.~Lazar, M.~Stöger-Pollach, S.~Löffler,
  A.~Steiger-Thirsfeld, G.~Van~Tendeloo,
  \href{http://link.aip.org/link/?APL/99/203109/1}{Atomic scale electron
  vortices for nanoresearch}, Appl. Phys. Lett. 99~(20) (2011) 203109.
\newblock \href {http://arxiv.org/abs/1405.7247} {\path{arXiv:1405.7247}},
  \href {http://dx.doi.org/10.1063/1.3662012} {\path{doi:10.1063/1.3662012}}.
\newline\urlprefix\url{http://link.aip.org/link/?APL/99/203109/1}

\bibitem{MaM_v18_i_p711}
H.~L. Xin, H.~Zheng, On-column 2p bound state with topological charge $\pm 1$
  excited by an atomic-size vortex beam in an aberration-corrected scanning
  transmission electron microscope, Microsc. Microanal. 18 (2012) 711--719.
\newblock \href {http://dx.doi.org/10.1017/S1431927612000499}
  {\path{doi:10.1017/S1431927612000499}}.

\bibitem{ACA_v68_i_p443}
S.~Löffler, P.~Schattschneider, Elastic propagation of fast electron vortices
  through crystals, Acta Crystallographica Section A 68~(4) (2012) 443 -- 447.
\newblock \href {http://arxiv.org/abs/1111.6050} {\path{arXiv:1111.6050}},
  \href {http://dx.doi.org/10.1107/S0108767312013189}
  {\path{doi:10.1107/S0108767312013189}}.

\bibitem{NC_v5_i_p5653}
K.~Müller, F.~F. Krause, A.~Béché, M.~Schowalter, V.~Galioit, S.~Löffler,
  J.~Verbeeck, J.~Zweck, P.~Schattschneider, A.~Rosenauer, Atomic electric
  fields revealed by a quantum mechanical approach to electron picodiffraction,
  Nature Communications 5 (2014) 5653.
\newblock \href {http://dx.doi.org/10.1038/ncomms6653}
  {\path{doi:10.1038/ncomms6653}}.

\bibitem{PRB_v76_i6_p60408}
J.~Rusz, O.~Eriksson, P.~Nov\'{a}k, P.~M. Oppeneer, Sum rules for electron
  energy loss near edge spectra, Phys. Rev. B 76~(6) (2007) 060408.
\newblock \href {http://dx.doi.org/10.1103/PhysRevB.76.060408}
  {\path{doi:10.1103/PhysRevB.76.060408}}.

\end{thebibliography}

\end{document}